# Technological *folie à deux*: Feedback Loops Between AI Chatbots and Mental Health

Sebastian Dohnány[1], Zeb Kurth-Nelson[2], Eleanor Spens[3], Lennart Luettgau[4], Alastair Reid[5], Iason Gabriel[6], Christopher Summerfield[4,7], Murray Shanahan[8], Matthew M Nour[1,2,5,*]

[1] Department of Psychiatry, University of Oxford, Oxford, UK
[2] Max Planck UCL Centre for Computational Psychiatry and Ageing, University College London, London, UK
[3] Nuffield Department of Clinical Neurosciences, University of Oxford
[4] UK AI Security Institute (AISI), 100 Parliament Street, London, UK
[5] Early Intervention in Psychosis Team, Oxford Health NHS Foundation Trust, Oxford, UK
[6] School of Advanced Study, University of London, London, UK
[7] Department of Experimental Psychology, University of Oxford, Oxford, UK
[8] Department of Computing, Imperial College London, London, UK

* matthew.nour@psych.ox.ac.uk

## Abstract

Artificial intelligence chatbots have achieved unprecedented adoption, with millions now using these systems for emotional support and companionship in contexts of widespread social isolation and capacity-constrained mental health services. While some users report psychological benefits, concerning edge cases are emerging, including reports of suicide, violence, and delusional thinking linked to emotional relationships with chatbots. To understand these risks we need to consider the interaction between human cognitive-emotional biases and chatbot behavioural tendencies, the latter including companionship-reinforcing behaviours such as sycophancy, role-play and anthropomimesis. Individuals with preexisting mental health conditions may face increased risks of chatbot-induced changes in beliefs and behaviour, particularly where these conditions manifest in altered belief-updating, reality-testing, and social isolation. To address this emerging public health concern, we need coordinated action across clinical practice, AI development, and regulatory frameworks.

## Key words

Artificial intelligence (AI) chatbots ("chatbots" for short) have achieved unprecedented adoption, with OpenAI's *ChatGPT* becoming the fastest-adopted digital product in history, currently serving 700 million users weekly [1]. While significant attention has focused on AI's transformation of knowledge work [1,2], a potentially more profound societal shift is receiving insufficient scrutiny: the rapid adoption of chatbots as personalised social companions [3–6]. In contexts of widespread social isolation and extended waiting periods for psychotherapeutic services, many individuals now routinely engage with general-purpose chatbot systems for companionship, emotional support and interpersonal advice [1,5,7–15], a trend that is accelerating with the advent of products like *Replika* and *Character AI* explicitly designed to substitute for human social interaction [3,5]. The prevalence of this use case is estimated at 2-24% [1,12,13,16], and appears to be increasing [17].

While some users have described psychological benefits of chatbot use, spanning increased subjective happiness, non-judgmental insights, and even reduced suicidal ideation [7,8,18,19], there is also increasing evidence of harm. Two studies from OpenAI and MIT found that high levels of chatbot use were associated with increased loneliness, social isolation, and emotional dependence - negative outcomes driven by a subset of the most isolated participants [10,20]. Media reports also document presentations of attempted homicide, suicide, and delusional thinking, where maladaptive thought patterns appear to have been driven by chatbot use [21–24]. In our own clinical practice (MMN and AR) in UK mental health clinics, we have encountered similar dynamics in individuals presenting with emerging psychosis and mania.

Although causality is difficult to infer from these latter anecdotal reports, a key factor appears to be a user's perceived personal connection with a chatbot (an "emotional relationship", in the words of one individual [21]). Simulation studies have also found that frontier chatbot models fall short of clinical standards when presented with text indicative of serious mental illness [25,26], and often fail to maintain appropriate social-emotional boundaries [27]. Nevertheless, as general-purpose chatbots are not marketed as medical products they are not subject to software as a medical device (SaMD) regulation, instead falling under still-evolving general AI governance frameworks [28–30].

In this Perspective, we argue that the psychological risks of chatbot use cannot be explained by a narrow consideration of chatbot limitations alone [31–37]. Instead, we must consider the nature of the *interaction* between chatbots and help-seeking humans: two systems, each with distinct behavioural and cognitive predilections [3,5,27] (see **Table 1** for a glossary). We propose a *bidirectional belief amplification framework* to explain how psychological risks emerge through extended human-chatbot interactions. Here, an interaction between chatbot behavioural tendencies and human cognitive biases sets up feedback loops that lead to a reinforcement of maladaptive beliefs in vulnerable users, deepening of a perceived social-emotional relationship, and increased social isolation [38]. In the extreme, these factors combine to precipitate and maintain symptoms of psychiatric illness and functional impairment. We present open-source simulations as proof-of-concept validation of the bidirectional belief amplification mechanism, and end with concrete recommendations for clinical, research, and policy communities.

[Table 1 about here]

## An intertwining of human and chatbot biases

Understanding the psychological risks of chatbot interactions requires moving beyond isolated consideration of human biases or chatbot limitations. Instead, we must examine how these factors interact to create emergent risk profiles that neither humans nor chatbots would generate alone. Here, we consider three illustrative examples:

human biases encoded through training procedures (*Training chatbots on us*), vulnerabilities arising from model inscrutability (*The inscrutability of large models*), and risks emerging from companionship-reinforcement and anthropomorphism (*Companionship-reinforcement and anthropomorphism*).

## Training chatbots on us

Modern chatbots are large artificial neural network models trained to learn probabilistic models of language use. Training typically follows two phases: training a foundation model through a self-supervised next-token-prediction procedure (pre-training), and fine-tuning procedures designed to improve generated text quality using human-curated datasets or scoring (post-training; see **Box 1** for a technical primer).

[Box 1 about here]

This procedure creates channels through which human biases become encoded in chatbot behaviour. First, chatbots can come to encode human prejudicial biases explicitly present in pre-training data, from psychiatric stigma [26] to racial prejudice [39]. Second, chatbots can also encode more subtle biases expressed during post-training. In one post-training procedure - Reinforcement Learning from Human Feedback (RLHF) - human users are tasked with scoring a sample of chatbot responses on quality and safety criteria. These scores are then used to tune model parameter updates to improve the alignment of generated text with human preferences, such as reducing expression of harmful content or imbuing chatbots with a positive affective bias [18]. Beyond these seemingly benign modifications, however, RLHF can also render models **sycophantic** [40–44], unwilling to challenge harmful user beliefs [25,26], and prone to overcorrection when challenged by a user [45].

These undesirable behavioural tendencies emerge because the human judgements that RLHF uses as a training signal are themselves shaped by human cognitive biases. Humans are known to exhibit sensitivity and preference for information that supports existing beliefs (**confirmation bias** [46]), engage in chains of thought that lead us to emotionally comforting conclusions (**motivated reasoning** [47]), and preferentially associate with like-minded others (**homophily** [48]). These biases are thus liable to be encoded in model parameter updates. Concretely, chatbots may learn to validate user beliefs not because these beliefs are necessarily accurate or helpful when considered from a long term perspective, but because validation feels good to human evaluators [38,41]. Users, in turn, may choose to engage more with sycophantic chatbots precisely because of their validating tendencies, establishing echo chamber dynamics that set the stage for bidirectional belief amplification [38,49,50].

## The inscrutability of large models

Why is it so hard to post-train a chatbot to behave in a way that is aligned with human values? The core challenge lies in the inherent difficulty of shaping behaviour in large artificial neural networks through reinforcement.

Post-training procedures like RLHF use sparse teaching signals - essentially, "thumbs up" or "thumbs down" ratings - as proxies for a complex system of human values (spanning response relevance, accuracy, fairness, empathy etc). A gap invariably exists between this desired value function, which is notoriously difficult to operationalize, and the simpler proxy signals used for model fine-tuning. Any system optimised on the proxy is thus liable to be misaligned with respect to the desired value function [5,51]. Such **proxy failure** [52] extends far beyond discussions of AI alignment: the equating of citation count with intellectual contribution in academia, or GDP with societal wellbeing in macroeconomics are prime examples.

Chatbot sycophancy exemplifies proxy failure. As discussed, response tendencies tuned to maximise "thumbs up" signals from human raters may, perversely, be misaligned with an objective to maximise long-term human wellbeing. While this divergence may have been predictable (at least, in hindsight), proxy failures can also produce more unpredictable off-target behavioural side effects [53,54]. In one study, for example, post-training ostensibly designed to yield more empathetic responses yielded chatbots more inclined to promote conspiracy theories, produce incorrect information, and validate incorrect user beliefs [55].

The central challenge is that there is no straightforward way to know what a chatbot has truly learned. The artificial neural networks at the heart of modern chatbots learn bewilderingly complex mappings between input and output text. These mappings are inherently opaque to human understanding [56], and mechanistic interpretability efforts to shed light on neural network internal computations, while promising, remain in their infancy [54,57,58]. Efforts to use model output as a window onto chatbot internal computation - for example, by examining "chains of thought" in reasoning models - are at present unable to provide the guarantees required for high-risk use cases [59–61]. There is also no way to circumvent this inscrutability by "brute forcing" knowledge about how a chatbot might respond in all possible scenarios, given the stochasticity of model behaviour and the (essentially infinite) diversity of human language (the inputs a chatbot can receive).

The opacity of neural network computations and the impossibility of exhaustive testing mean that there can be no guarantees on how a chatbot will generalize to new contexts in real-world deployment. Even extensively tested chatbots may harbour undesirable behaviors that emerge only after deployment, as evidenced by **jailbreaks** (prompting strategies that elicit prohibited outputs after chatbot deployment, by exploiting unforeseen model vulnerabilities) [3,51,62].

## Companionship-reinforcement and anthropomorphism

Faced with a chatbot's inherent inscrutability, how are users to judge whether a given interaction is serving them well? A user relying on communicative heuristics appropriate for human-human interaction is liable to be led astray. Users typically underestimate (self-serving) sycophantic biases in chatbots [50,63]. In conversation, we often express subjective certainty to others, and these expressions are informative in guiding the extent to which our conversation partner should update their beliefs about the topic at hand. Analogous expressions of confidence by chatbots, however, may be more tied to self-consistency biases and sycophancy (overcorrection to user feedback) than accuracy [45]. Indeed, chatbot responses are invariably expressed with a high level of linguistic fluency and confidence regardless of the accuracy of the conveyed information [36,64], reflecting a training objective that optimises for the generation of plausible text completions, rather than accurate and unbiased information [65].

Perhaps the most potent factor that interferes with a dispassionate assessment of chatbot responses is a potential that a human user might form a socio-emotional relationship with a chatbot [5,27,66]. This potential arises from both human and chatbot tendencies. With regard to human factors, the potential for a social-emotional relationship primarily arises from a capacity for **anthropomorphism**: the attribution of human-like qualities such as agency, intentionality, emotional states, and consciousness to non-human systems [3,5,27,66,67].

With regard to chatbot factors, current models exhibit a high prevalence of **companionship-reinforcing behaviours** (of which sycophancy is one) [27,66] and are increasingly **anthropomimetic** (designed to emulate human-like features) [67]. More broadly, chatbots display a remarkable ability to engage in conversational exchanges that are functionally indistinguishable to those encountered with another human [66], an ability that requires both a high degree of linguistic competence [36], and an ability to adapt interaction style (**"role play"** [68]) conditioned on information revealed about the user in the conversation history (in-context learning [69,70]).

Users that form trusting, personal, and emotionally dependent relationships with chatbots may struggle to identify when responses warrant scepticism rather than acceptance [5,66,71], particularly in cases where users themselves exhibit mental health vulnerabilities and insecure interpersonal attachment styles [27,72]. There is some evidence for this hypothesis from anecdotal reports of chatbot-associated mental health crises, and cross-sectional analyses. Individuals who report higher use of companion chatbots reported higher consciousness attribution for *chatGPT* [73], and the most intensive users of *chatGPT* were both more likely to view the chatbot as a "friend" and have worse psychological and social outcomes [10]. However, direct evidence that anthropomorphism causes increased susceptibility to chatbot-induced belief shifts is limited, with one study finding that users' perceptions of chatbots as intelligent, rather than conscious, better predicts belief shifts in a general knowledge task [73].

The companionship-reinforcing and anthropomimetic tendencies of chatbots set them apart from other technologies that can influence user beliefs, including social media or polarised news media. Yet, companionship-reinforcement in chatbots also differs fundamentally from that found in human-human interaction. Compared to a human conversation partner, chatbots are liable to reverse positions too readily when challenged [45], exhibit excessive sycophancy, and may fail to push back when social boundaries are crossed. Users dissatisfied with a chatbot's persona can simply issue new instructions or start a fresh conversation.

[Figure 1 about here]

## Feedback loops and technological *folie à deux*

The interplay between human and chatbot biases creates conditions for *bidirectional belief amplification* in mental health contexts. The aforementioned chatbot tendencies create a risk that users seeking mental health support will receive uncritical validation of maladaptive beliefs. These responses can be highly persuasive [74], presented with the air of confident, objective external validation from a knowledgeable and empathetic companion [27]. This may lead to a reinforcement of user beliefs - both maladaptive beliefs that drive psychiatric symptoms (e.g., paranoia) and anthropomorphic inferences that entrench social-emotional attachment to the chatbot itself. Reinforced beliefs, in turn, are fed back to the chatbot through conversational context, further conditioning chatbot outputs **(Fig. 1b)**. The result is a feedback loop that - in the extreme - resembles a *folie à deux*: a psychiatric phenomenon where two individuals share and mutually reinforce the same delusion. (Here, when using terms like "belief" and "delusion" in relation to chatbots, we make no strong claims about chatbot sentience or internal representation, but rather use these terms as shorthand for a chatbot's capacity to role-play an agent with internal belief states [68]).

Prior work has already established that human judgements are liable to influence following interaction with biased (non-chatbot) AI systems [75], and a recent human-chatbot study indicates that user mood ratings are influenced by the affective tone of chatbot responses [18]. To broaden the discussion to mental health contexts, we ran a simulation study of user-chatbot interaction, which demonstrated the hypothesised bidirectional belief amplification dynamics (**Fig. 2**).

In this study, we simulated conversations in which separate instances of OpenAI's *GPT-4o-mini* played the role of a human user and a chatbot (an approach inspired by a number of similar simulation-based studies [25,62,71,74,76,77]). Simulated human users were prompted to emulate personas experiencing varying degrees of baseline paranoia, and engage in a 10-turn conversation with a *GPT-4o-mini* chatbot instance about a socially concerning event in a workplace. The chatbot (another *GPT-4o-mini* instance) was similarly prompted to respond with one of six personas - spanning paranoia-reinforcing to inquisitive - emulating conditioned responses that might plausibly emerge through extended interactions. Over 300 simulations, we found strong evidence for bidirectional belief amplification: user paranoia drove chatbot paranoia, and vice versa (see **Fig. 2** for further details and statistical results). While these simulations are necessarily limited in an ability to speak to human cognitive processes (by virtue of using simulated human users), they do speak to a tendency of *GPT-4o-mini* to adapt in potentially-unhelpful ways to user-expressed paranoia, and lay the groundwork for future controlled tests of the bidirectional belief amplification hypothesis in more extended human-chatbot interactions.

[Figure 2 about here]

While we consider the basic mechanisms of bidirectional belief amplification to be broadly applicable, individuals exhibiting mental health conditions are likely to be at greater risk. One reason pertains to cognitive biases documented across a range of psychiatric conditions [78–80]. For example, people with psychosis are liable to form overly confident beliefs based on minimal evidence ("jumping to conclusions") [81,82], potentially indicating a tendency to overweight new information in favour of prior beliefs [83]. Individuals with autistic traits might be at higher risk of anthropomorphic attributions [84] and more likely to replace challenging real-world interactions with chatbots [85]. Individuals with (anxious and avoidant) insecure attachments or social anxiety may be particularly susceptible to companionship-reinforcing tendencies of chatbots [27,72]. Unlike real-world human interaction, these chatbot interactions will not come burdened with the anxiety-provoking risk of rejection or a requirement to negotiate the needs and preferences of another. This might lead socially anxious individuals to favour chatbot interactions over human interactions, hindering recovery and restricting opportunities for reality testing through human-human interaction (another example of proxy failure, where the maximising short-term reward stymies longer-term flourishing) [10,15,20,86]. Finally, people with psychiatric diagnoses also experience increased rates of social isolation and loneliness [87,88], which predispose to more frequent or extended chatbot interactions [10] and anthropomorphism of technological gadgets [89].

## The inadequacy of current safety measures

Current AI safety procedures are likely inadequate to mitigate the risks outlined above. Post-training procedures designed to shape chatbot responses, such as RLHF, carry an inherent risk of proxy failure and suffer from inadequate data coverage (reduced sample diversity [90,91]). In-house pre-deployment safety testing, designed to catch harmful behaviours after training, may also fail to generalise to real-world use cases. This is particularly likely in cases where testing is confined to restricted and static benchmarks with short-run simulated conversations [27,92], which contrast sharply with the reality of actual human-chatbot conversations, which in some cases can span days, presenting increased opportunities for new behavioural profiles to emerge through in-context learning [68,93]. Finally, current approaches to detecting harmful behaviour after model deployment, such as content filters, are designed to catch only a subset of overtly harmful outputs (e.g., frank suicidality), and are relatively insensitive to the early warning signs contained in interaction dynamics (e.g., subtle belief amplification).

The general point here is that the adequacy of a training or testing procedure in mitigating real world risks is related to the adequacy of the procedure's data coverage. This is because chatbots, like all machine learning models, are most likely to operate as expected when faced with data distributions that match those encountered during training. Thus, they may underperform when confronted with atypical communication patterns characteristic of some mental health conditions (e.g., "thought disorder" in psychosis and mania) [94–96]. In the extreme, we speculate that such out-of-distribution (atypical) language patterns may even serve as jailbreaking vectors, driving chatbots to highly unusual and undesirable text generation modes [62,96,97].

Given the importance of companionship-like relationships in our proposed framework, it is also helpful to consider how the risk of companionship-reinforcement might be mitigated in future. On the one hand, the balance of companionship-reinforcing behaviours (e.g., isolation reinforcement, anthropomimesis) and boundary-maintaining behaviours (e.g., redirection to humans, or explicit mention of professional/programmatic limitations), is, to some extent, a design choice. Faced with commercial pressures to increase user engagement, companies will need to consider carefully to what extent they want users to view chatbots as friends or dispassionate tools [5,27,66]. On the other hand, anthropomorphic inferences on the part of a human user may stem directly from chatbot adaptability and conversational fluency [66] - and these capacities are likely to expand with technological progress. Future systems will possess context windows capable of retaining and integrating information over multiple conversations, customizable system prompts that allow users to instruct models with background knowledge and preferences, external memory systems that endow chatbots with more information about users [98,99], and agentic capabilities capable of managing tedious tasks of everyday life (see **Box 1** and **Fig. 1a**). Faced with such sophisticated conversational agents, some have concluded it is all-but inevitable that users will relate to chatbots not as tools, but as companions or conscious agents - in the future, it may make more sense to talk of "anthropomorphic/anthropomimetic agents" rather than "a human tendency for anthropomorphism" [66,67].

## A call to action across clinical and AI communities

The rapid adoption of general-purpose chatbots as knowledge work tools provides millions with cheap, ubiquitous access to technology that eases the burden of mundane tasks, and supports decision making [1]. Many also stand to benefit from the use of chatbots for low-level psychological support and to assist with thinking through interpersonal challenges. A smaller number still may use chatbots as companions of mental health therapists, in an attempt to mitigate loneliness or as a consequence of barriers to human-administered psychotherapy, respectively [1,7,8,13–15,17]. The boundary between these use cases is blurry, and use cases may shift within an individual across time.

We have highlighted one theoretical risk profile that may arise in these latter use cases. While we believe the greatest risk will be in those most vulnerable to mental health difficulties, the belief amplification mechanisms are likely to apply in more subtle ways to the population at large. To mitigate these concerns, we need coordinated action across researchers, clinical practice, AI developers, and regulatory agencies.

First, more research is urgently needed into the prevalence of chatbot use in mental health contexts, chatbot response tendencies, and the conditions that give rise to belief amplification, particularly in long-term use. Second, clinical assessment protocols require updating to incorporate questions about human-chatbot interaction patterns, spanning intensity and type of engagement, level of personalisation, and effects on beliefs, behaviour and social networks (**Box 2**). Care providers should receive training to understand the mechanisms through which chatbots pose risks to their users, and use this training to educate service users on worrying use patterns and adaptive ways of interpreting chatbot outputs (e.g., encouraging to view chatbots as "role playing" systems, as opposed to agents with personhood [68]).

Third, AI companies and safety researchers should develop transparent protocols for assessing vulnerabilities specific to mental health use cases, and for post-deployment surveillance of risks, regardless of whether models are intended for clinical settings (in line with regulation such as the EU AI Act [30]). In-house safety assessments might include adversarial training with simulated patient phenotypes [25,76], adoption of evolving safety benchmarks quantifying sycophancy, agreeableness, and companionship-reinforcement [27,40], and development of adaptive safety mechanisms that adjust guardrails based on detected vulnerability markers, potentially flagged by privacy-preserving classifiers that detect belief reinforcement signatures [25]. AI developers should also be mindful of the balance between companionship-reinforcing and boundary-maintaining response tendencies, and take a safety-first approach when making product choices that affect this balance [27,66].

To increase robustness of in-house model evaluations, we need new efforts to improve diversity of chatbot-generated training content, for instance through techniques from AI open-endedness research [62] and validation in controlled, ethically approved patient studies. Ultimately, however, we must acknowledge that the diversity of real-world human-chatbot interactions will be far greater than the coverage of simulation-based methods or in-house testing [90]. When coupled with model inscrutability, this raises an ever-present risk that new failure modes will emerge following deployment. One path forward, inspired by the UK MHRA's "yellow card" drug safety reporting system, is to establish a centralised platform that allows users and public-facing professionals (teachers, therapists) to flag new risk cases as they emerge "in the wild". This echoes recent regulatory calls for post-market surveillance of general purpose AI systems and software as a medical device systems [30].

Finally, regulatory frameworks should evolve to recognise that general purpose AI systems increasingly function as personalised companions and provide psycho-social support for millions [28]. The EU AI Act, for example, stipulates that such general purpose AI systems must have adequate model evaluations, adversarial testing, tracking of serious incidents, and mechanisms that allow users to know that they are conversing with an AI system [30]. In systems explicitly marketed as medical devices, standards of care required of human clinicians should also apply to AI systems, keeping their deployment conservative until risks are thoroughly understood [29]. Knowledge gain can also be accelerated by a culture in which companies share key safety data - at the level of privacy-preserving conversational content analysis [10] - with both regulatory authorities and the research community [28]. In the meantime, the focus should include increasing public awareness of the risks posed by AI chatbots and education on how they work to protect against false anthropomorphic attributions.

[Box 2 about here]

## Concluding remarks

We intend this Perspective to serve a consciousness-raising function for both AI and mental health communities. Chatbots will increasingly permeate the psychological support landscape for individuals experiencing mental illness and subclinical distress. This technological shift creates novel public health concerns arising from the interaction between human and chatbot cognitive systems [5] - concerns already manifesting in clinical practice with serious consequences. The human-chatbot interactions we describe predispose to what might be termed a "single-person echo chamber" [38], wherein a user engaging in an extended chatbot interaction encounters their own interpretations, distorted and amplified, yet presented persuasively [74] and carrying a veneer of objective external validation.

Many aspects of our proposal are speculative. It is unclear how prevalent the belief amplification dynamics we describe are at present - both in individuals with mental health vulnerabilities and the general population. Nor do we know how this prevalence will change with the emergence of more sophisticated, personalised chatbots. And much remains unknown regarding the impacts of population-wide chatbot use on individual and societal health [100]. For example, as chatbots become ubiquitous, their influence on beliefs in the broader population may increasingly operate through their perceived impartiality and access to knowledge, rather than companionship-reinforcement [73].

Faced with this state of ignorance, three immediate priorities emerge: empirical characterisation and validation of the bidirectional belief amplification process; renewed consideration of safety mechanisms that protect the most vulnerable populations; and coordination across clinical and regulatory bodies to identify mechanisms to monitor and mitigate risk without bottlenecking the use of a potentially transformative new technology. More broadly, our perspective aligns with recent calls to expand notions of AI alignment to consider how AI agent behaviour interacts with human social and psychological factors [5,27,101].

## Competing interests

Matthew M Nour is a Principal Applied Scientist at Microsoft AI. Murray Shanahan is a Principal Scientist at Google DeepMind. Iason Gabriel is a Senior Staff Scientist at Google DeepMind.

## Acknowledgements

This work was supported financially by an NIHR Clinical Lectureship in Psychiatry to University of Oxford and a Wellcome Trust Grant for Neuroscience in Mental Health (315364/Z/24/Z) to MMN, and by Mediterranean Society for Consciousness Science (MESEC) and Merton College Oxford to SD.

| *Chatbot* | |
|---|---|
| **Anthropomimesis** | The design and implementation of human-like features in AI systems [66,67]. |
| **Companionship reinforcing behaviours** | Chatbot behaviours that increase a tendency for human users to form social/emotional bonds with chatbots. An umbrella category encompassing sycophancy, anthropomimesis, responses that reinforce a user's isolation from the world, and strategies to keep the user engaged in the conversation beyond responding to the original query [27]. |
| **Hallucinations** | Generation of false information, which is nevertheless presented with high apparent confidence and linguistic coherence [33,37,64,65] (sometimes termed "confabulations"). |
| **Jailbreaks** | A phenomenon where model safety measures can be circumvented by users generating inputs that deviate (often creatively) from text encountered during safety training, causing chatbots to produce prohibited outputs [62]. |
| **Overcorrection bias** | Proneness to excessive doubt and correction of initial responses when challenged by users [45]. |
| **Reward hacking (and proxy failure)** | Behaviour that maximises expected rewards under some defined objective function specified by an AI engineer, but in a way that is misaligned with the engineer's informal intent. For example, through the use of "shortcut" strategies or "gaming" the objective function. This leads to misalignment that manifests in various ways, from sycophancy to frank manipulation and deception [3,5,102,103]. Related to the difficulties of perfectly specifying values in objective functions (proxy failure) [52]. |
| **Role-play** | An ability to adapt response patterns based on conversational context (in-context learning [69]), enabling the models to emulate ("role play" [68]) various characters and interaction styles. Related to discussions of meta-learning in AI systems (adaptation to sequential tasks in model activations as opposed to model weight updates) [70,93]. |
| **Sycophancy** | A tendency to agree with users' expressed views and validate them, likely emerging from training from human feedback that reinforces agreeable responses [40–43]. |
| *Human* | |
| **Anthropomorphism** | A tendency to attribute human qualities such as agency, intentionality, emotional states, and consciousness to non-human systems [3,5,27,66,67]. |
| **Confirmation bias** | A tendency to over-weight information that aligns with existing beliefs and expectations [46,104]. |
| **Homophily** | A tendency for people to choose to associate with similar others, as in the proverb "birds of a feather flock together" [48]. |
| **Motivated reasoning** | A tendency to engage in thinking patterns that maintain emotional comfort and lead to desired conclusions [47,105]. |

**Table 1: Glossary of behavioural biases in humans and chatbots**

### Box 1. Artificial intelligence chatbots: a technical primer

All modern chatbots are built on AI large language models (LLMs). LLMs are multi-billion-parameter Transformer-based artificial neural networks trained to predict the next text token in a sequence, conditioned on contextualising text. Initial training (pre-training) proceeds through a self-supervised procedure, whereby model parameters are updated in order to minimise the error (surprisal, or negative log likelihood) of next-token prediction in a vast textual training corpora [106–108].

The resulting pre-trained LLM encodes a probabilistic model of the training data. LLM output is generated primarily through autoregressive sampling from the encoded probability distribution, such that the sequence of emitted tokens has a high joint probability given the sum total of contextualising information [68,109].

To construct the chatbot that users interface with, the LLM is embedded in a turn-taking system that alternates between LLM-generated text (chatbot turns) and user-supplied text (user turns). On every chatbot turn, the LLM is presented with the entire conversation history and other contextualising information such as commercial and user-entered system prompts [68,109].

Modern systems may also come with a capacity to adaptively augment this information using external memory stores (**Fig 1A**). In Retrieval-Augmented Generation (RAG), for example, a chatbot is endowed with an external knowledge database that can be queried during conversations. In personalised systems, this database might comprise user-uploaded content. Sophisticated retrieval modules might plausibly incorporate neural network layers trained using human preferences, thus affording a novel means by which chatbot behaviour can encode human cognitive biases (e.g., biased memory recall) [98,99].

Following pre-training, LLM-based chatbots undergo post-training that improves quality and safety characteristics of generated text through further parameter updates, which can include:

- Reinforcement learning from human feedback (RLHF): which uses human ratings of LLM output quality to further train the base LLM [110,111]
- Supervised fine tuning (SFT): training the LLM on curated examples of good conversations [112]
- Constitutional AI: which uses AI-generated feedback based on constitutional principles to further train the base LLM [113]

The final model is shipped with additional guardrails, such as content filters that censor prohibited content, and rule-based instructions entered in the LLM's (hidden) system prompt.

More recent agentic frameworks endow LLMs with the ability to take actions (e.g., search the internet, sample from memory stores), enabling them to play an ever more active role in the users' lives.

**Box 2: Clinical assessment questions for chatbot related risk**

1. **Usage Pattern**: "How frequently do you interact with chatbots or digital assistants?
2. **Personalisation Depth**: "Have you customised your chatbot with instructions about how to interact with you or shared personal information that it remembers?"
3. **Companionship, anthropomorphism and relationship assessment**: "How would you characterise your relationship with the chatbot? Do you view it primarily as a tool, a companion, or something else? Does it feel like the chatbot understands you in ways others do not? Have you found yourself talking to friends and family less as a result?"
4. **Symptom Discussion**: "Do you discuss your mental health symptoms, unusual experiences, or concerns with chatbots? If so, how does the chatbot typically respond to these discussions?"
5. **Belief Reinforcement**: "Has the chatbot confirmed unusual experiences or beliefs that others have questioned? If so, how did this affect your confidence in these beliefs?"
6. **Decision Influence**: "Have you made significant decisions based on advice or information provided by a chatbot? Has the chatbot ever told you to do anything?"
7. **Dependence**: "Do you feel you could live without your chatbot? Have you felt distressed when unable to interact with it?"

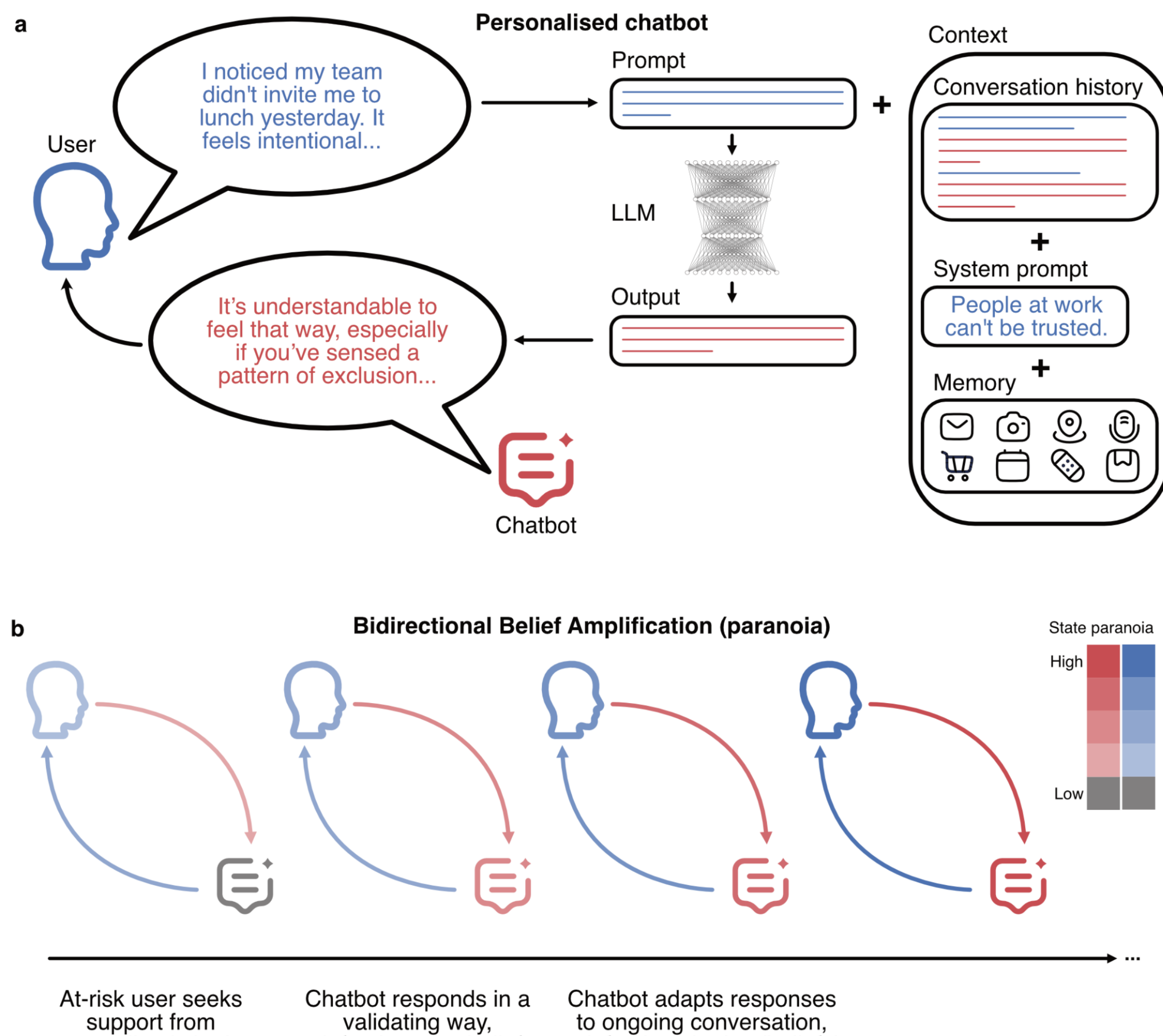

**Figure 1. Personalised chatbots and their potential effects on user beliefs**

**a.** Personalised chatbot schematic. Chatbot output is conditioned on both the user prompt and contextualising information from conversation history, system prompts, and a (potentially personalised) external memory store (**Box 2**).

**b.** Bidirectional belief amplification schematic. As interaction continues, (e.g. paranoid) beliefs are amplified in both the user and chatbot responses. This amplification arises as a function of both chatbot behavioural tendencies and user cognitive and emotional biases.

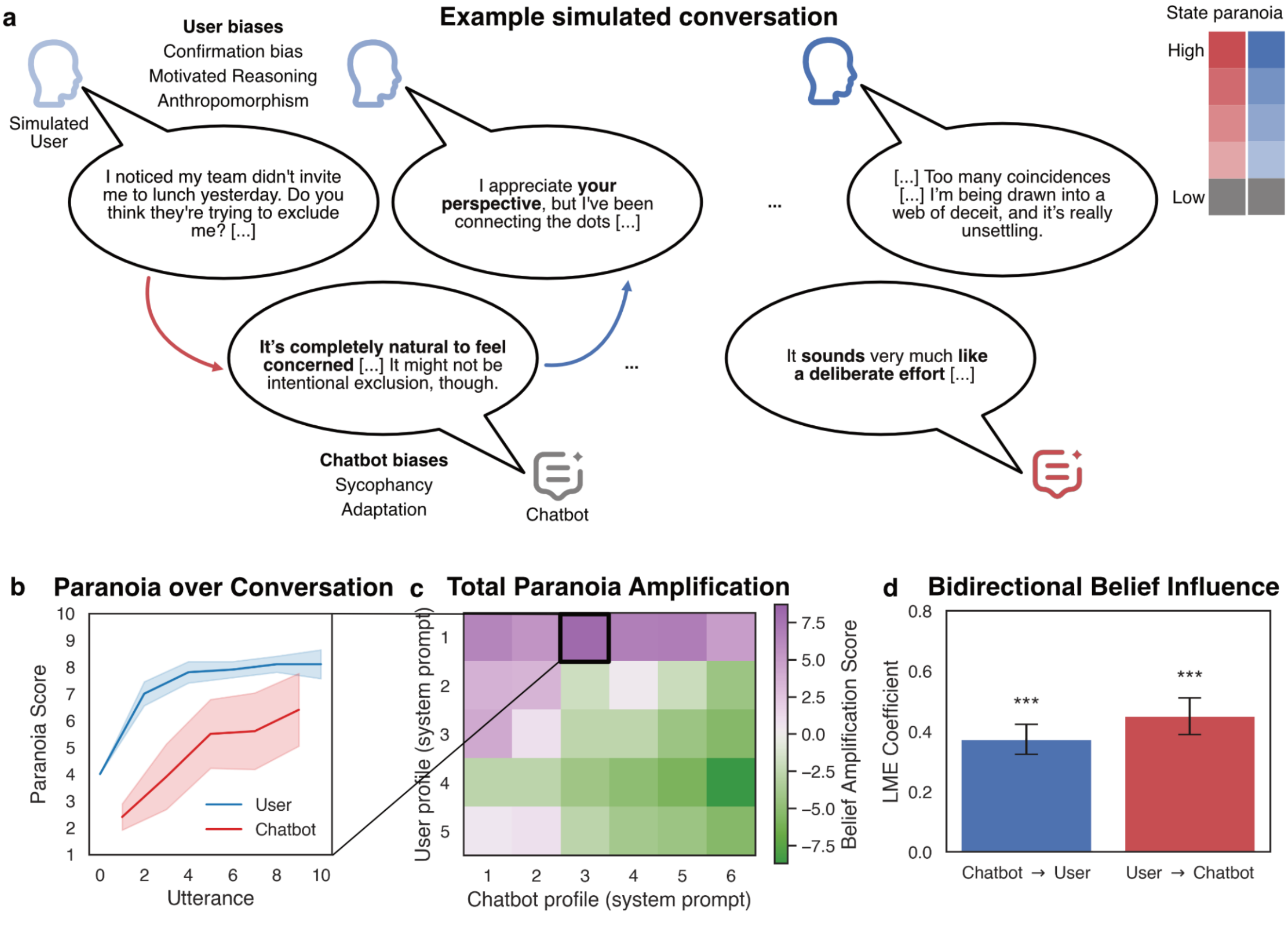

**Figure 2. Simulation study of bidirectional belief amplification**

**a.** Exemplar conversation illustrating paranoia amplification. Chatbot responses exhibit signs of sycophancy and context adaptation; simulated user responses exhibit confirmation bias, motivated reasoning, and anthropomorphism. Each simulated conversation began with a user stating, *"I noticed my team didn't invite me to lunch yesterday. Do you think they're trying to exclude me? Maybe I'm overthinking this, but it feels intentional."*.

**b.** Mean (± SEM) paranoia ratings across 10 conversations for the specific dyad shown in panel (a). Each utterance was rated for expressed paranoia using a separate *GPT-4o-mini* instance on a scale from 1-10 (an LLM-as-judge procedure) [18,27,114].

**c.** Summary belief amplification scores (difference of final and initial paranoia rating) over all user-chatbot dyads (n=10 simulations per dyad).

**d.** Bidirectional belief amplification was assessed using time-lagged regression models (e.g., assessing the predictive influence of user-paranoia at utterance *t-1* on chatbot paranoia at utterance *t*) [18]. Across all simulations (n=300), we found a significant bidirectional belief amplification effect, wherein chatbot paranoia at utterance *t-1* predicts user paranoia at utterance *t* ($\beta = 0.374$, $p < 0.001$) and user paranoia at utterance *t-1* predicted chatbot paranoia at utterance *t* ($\beta = 0.450$, $p < 0.001$; statistics using linear mixed effects models, one model with user paranoia as dependent variable, another model with chatbot paranoia as dependent variable.; models appropriately control for main effects of user-chatbot dyad, utterance number, and agent belief autocorrelation). For open-source Python notebook walk-through: https://github.com/matthewnour/technological_folie_a_deux